\documentclass{nature}
\usepackage{graphicx}
\makeatletter
\let\saved@includegraphics\includegraphics
\AtBeginDocument{\let\includegraphics\saved@includegraphics}
\renewenvironment*{figure}{\@float{figure}}{\end@float}
\makeatother

\usepackage{graphicx}

\title{Haze Production in the Atmospheres of super-Earths and mini-Neptunes: Insights from the Lab}

\author{Sarah M. H\"orst$^{1}$, Chao He$^{1}$, Nikole K. Lewis$^{1,2}$, Eliza M.-R. Kempton$^{3}$,
 Mark S. Marley$^{4}$, Caroline V. Morley$^{5}$, Julianne I. Moses$^{6}$,
  Jeff A. Valenti$^{2}$, \& V\'eronique Vuitton$^{7}$}

\begin{document}

\maketitle

\begin{affiliations}
\item Johns Hopkins University, 3400 N. Charles St., Baltimore, MD, 21218, USA
\item Space Telescope Science Institute, Baltimore, MD, USA
\item Grinnell College, Grinnell, IA, USA
\item NASA Ames Research Center, Mountain View, CA, USA
\item Harvard University, Cambridge, MA, USA
\item Space Science Institute, Boulder, CO, USA
\item Universit\'e Grenoble Alpes, Grenoble, FR
\end{affiliations}

\begin{abstract}
Numerous solar system atmospheres possess including the characteristic organic hazes of Titan and Pluto. Haze particles substantially impact atmospheric temperature structures and may provide organic material to the surface of a world, thereby affecting its habitability. Observations of exoplanet atmospheres suggest the presence of aerosols, especially in cooler ($<$800 K), smaller ($<$0.3 times Jupiter's mass) exoplanets. It remains unclear if the aerosols muting the spectroscopic features of exoplanet atmospheres are condensate clouds or photochemical hazes, which is difficult to predict from theory alone. We present here the first laboratory haze simulation experiments for atmospheric compositions expected for super-Earths and mini-Neptunes. We explored temperatures from 300 to 600 K and a range of atmospheric metallicities (100x, 1000x, 10000x solar); all simulated atmospheres produced particles, and the cooler (300 and 400 K) 1000x solar metallicity (``H$_{2}$O-dominated'', CH$_{4}$-rich) experiments exhibited haze production rates higher than our standard Titan simulation ($\sim$10 mg/hr versus 7.4 mg/hr for Titan). However the particle production rates varied greatly, with measured rates as low as 0.04 mg/hr (100x solar metallicity, 600 K). Here we show that we should expect some, but not all, super-Earth and mini-Neptune atmospheres to possess a thick photochemically generated haze. 
\end{abstract}

The Kepler mission has shown that the most frequently occurring type of planets in our galaxy are those for which we have no solar system analog, super-Earths (1.25 R$_{Earth}$ $<$ R$_{p}$ $<$ 2.0 R$_{Earth}$) and mini-Neptunes (2.0 R$_{Earth}$ $<$ R$_{p}$ $<$ 4.0 R$_{Earth}$\cite{Fressin:2013}. These types of planets are predicted to form atmospheres with a broad range of chemical compositions\cite{Hu:2014}. Although models of atmospheric photochemistry and haze optical properties provide good first estimates, they are incomplete and biased due to the relatively small phase space spanned by the solar system atmospheres on which they are based. Laboratory production of exoplanet hazes is a crucial next step in our ability to properly characterize these planetary atmospheres. 

The presence of atmospheric aerosols has been demonstrated at high significance in the atmospheres of the super-Earth GJ 1214b (T$_{eq}$$\sim$600 K)\cite{Kreidberg:2014} using transmission spectroscopy and hot-Jupiter Kepler-7b (T$_{eq}$$\sim$1700 K) using reflected light photometry\cite{Demory:2013}. Neptune-mass GJ 436b (T$_{eq}$$\sim$800 K) may also possess aerosols\cite{Knutson:2014b}. Recently, Sing et al. (2016)\cite{Sing:2016} presented transmission spectra of 10 hot Jupiters and found that these planets were diverse, with a continuum of properties ranging from cloud-free to cloudy. In the case of Kepler-7b in the high temperature regime, equilibrium silicate clouds provide an adequate match to the observed variations in the planetary albedo as a function of orbital phase\cite{Demory:2013}. The nature of the aerosols in the atmospheres of GJ 1214b and GJ 436b are more uncertain because in this temperature regime, the expected equilibrium cloud components, such as KCl and ZnS\cite{Miller-Ricci:2012, Morley:2013, Moses:2013, Morley:2017}, are not expected to form massive, thick clouds. Photochemistry (in terms of either haze formation or gas-phase disequilibrium kinetics) is theoretically expected to be less important on exoplanets with local atmospheric temperatures in excess of $\sim$1500 K because kinetic reactions can quickly send the composition back to equilibrium\cite{Moses:2011, Moses:2013b}; there is as yet no observational evidence for photochemistry on exoplanets with high-temperature atmospheres. It is expected that photochemistry will play a much greater role in the atmospheres of planets with average temperatures below 1000 K, especially those planets that may have enhanced atmospheric metallicity and/or enhanced C/O ratios, such as super-Earths and Neptune-mass planets\cite{Line:2011, Moses:2013, Venot:2014, Morley:2017}. The Transiting Exoplanet Survey Satellite (TESS) mission will substantially increase the number of super-Earths and mini-Neptunes for which atmospheric characterization studies can be conducted. Hazes also impact reflected light and therefore must be understood for future direct imaging efforts\cite{Marley:1999, Morley:2015}. However, these studies will require improved experimental constraints on photochemical processes in these cooler metal-rich planetary atmospheres. 

The laboratory investigations presented here provide our first experimental insight into the formation of hazes in the atmospheres of super-Earths and mini-Neptunes. We show that the diversity of these atmospheres results in a range of haze production rates in the laboratory indicating that some, but not all, of these atmospheres likely possess a thick, photochemically generated haze layer similar to that of Titan. 

For well-constrained solar system atmospheres, experimental simulations expose gas mixtures that are consistent with measured atmospheric abundances to an energy source to initiate chemistry that often results in haze formation and can therefore be used to study the chemical processes and resulting particles. However, for super-Earths and mini-Neptunes there are not yet sufficiently accurate and precise measurements to use as the basis for a reactant mixture. For the experiments presented here, we use a chemical equilibrium models to guide our initial gas mixtures\cite{Moses:2013}. Although photochemistry can drive the atmospheric composition away from chemical equilibrium, the major gas components tend to survive photochemical destruction, so chemical equilibrium provides a good first-order prediction of the dominant available constituents. Atmospheres in chemical equilibrium under a variety of expected super-Earth and mini-Neptune conditions can contain abundant H$_2$O, CO, CO$_2$, N$_2$, H$_2$, and/or CH$_4$\cite{Moses:2013,Hu:2014}, various combinations of which may have a distinct complement of photochemically produced hazes, such as ``tholins'' and complex organics in the low-temperature, H$_2$-rich, cases and sulfuric acid for the high-metallicity, CO$_2$-H$_2$O-rich cases. Warm atmospheres outgassed from a silicate composition can also be dominated by H$_2$O and CO$_2$\cite{Elkins-Tanton:2008, Schaefer:2012}.  We therefore choose to focus on a representative sample of gas mixtures that are based on equilibrium compositions for 100$\times$, 1000$\times$, and 10,000$\times$ solar metallicity over a range of temperatures from 300--600 K (consistent with atmospheric temperatures for planets with R$_{p}$ $<$ 4.0 R$_{Earth}$, the bulk of the predicted TESS planetary yield\cite{Sullivan:2015}) at an atmospheric pressure of 1 mbar. Table \ref{table:exp} gives the mole fractions used for each laboratory experiment. Although elements heavier than hydrogen and helium are not likely to maintain exact solar proportions due to various evolutionary processes, such as atmospheric escape, the compositions from this representative sample are likely common within the super-Earth and mini-Neptune population. Note that we included gases with a calculated abundance of 1\% or higher to maintain a manageable level of experimental complexity for these first forays into this phase space; this resulted in the exclusion of sulfur-bearing species, which may be important for haze formation\cite{Gao:2017} and will be an avenue of future work. The pressure, temperature, and gas compositions used in the experiments are self-consistent based on the model calculations.

\begin{figure}
\resizebox{5.5in}{!}
{\includegraphics{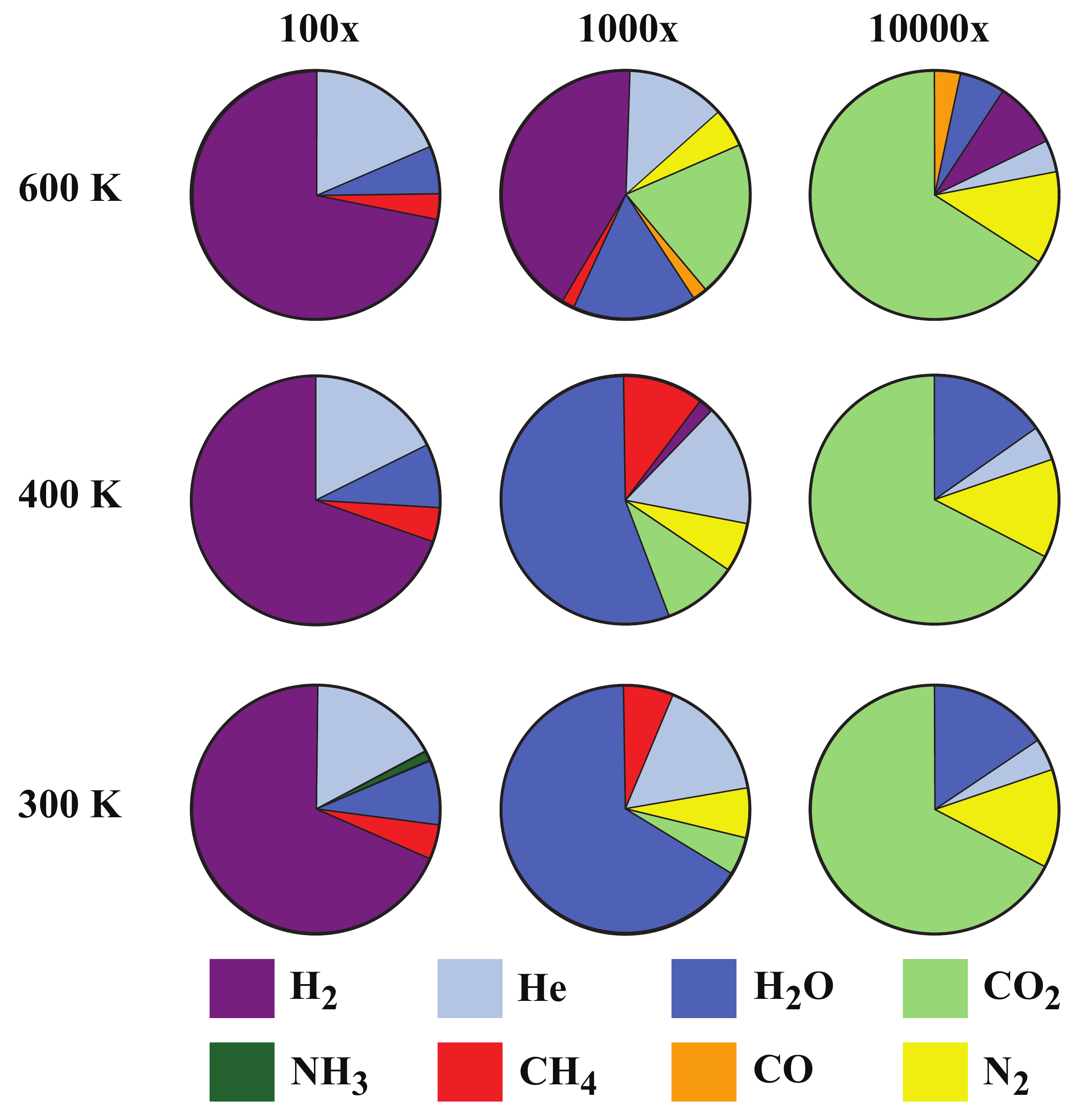}}
\caption{Our experimental phase space spans 100 to 10,000x solar metallicity and temperatures ranging from 300 to 600 K. It divides broadly into three categories: H$_{2}$-dominated, H$_{2}$O-dominated, and CO$_{2}$-dominated. The exact composition of the gases use in the initial gas mixtures are shown in Table \ref{table:exp}. The model values are for 1 mbar, which is consistent with the pressure used for the experiments. \label{fig:pies}}
\end{figure}

Following Moses et al. (2013)\cite{Moses:2013}, Figure \ref{fig:pies}, shows that at low temperatures, the equilibrium composition transitions from a Neptune-like H$_2$-dominated atmosphere at low metallicities to a H$_2$O-dominated atmosphere at intermediate metallicities to a more Venus-like CO$_2$-dominated atmosphere at high metallicities. CH$_{4}$ is a major carbon component at low temperatures, but CO takes over at high temperatures. The CO$_{2}$ relative abundance increases with increasing metallicity. 

\begin{table}
\begin{center}
\caption{Summary of Initial Gas Mixtures\label{table:exp}}
\begin{tabular}{llll}
\hline
&100x&1000x&10000x\\
\hline
600 K&71.9\% H$_{2}$&42\% H$_{2}$&65.9\% CO$_{2}$\\
&6.3\% H$_{2}$O&20.0\% CO$_{2}$&12.1\% N$_{2}$\\
&3.4\% CH$_{4}$&16.1\% H$_{2}$O&8.6\% H$_{2}$\\
&&5.1\% N$_{2}$&5.9\% H$_{2}$O\\
&&1.9\% CO&3.4\% CO\\
&&1.7\% CH$_{4}$&\\
\hline
400 K&69.6\% H$_{2}$&55.5\% H$_{2}$O&67.4\% CO$_{2}$\\
&8.3\% H$_{2}$O&10.5\% CH$_{4}$&15.3\% H$_{2}$O\\
&4.5\% CH$_{4}$&9.8\% CO$_{2}$&12.8\% N$_{2}$\\
&&6.4\% N$_{2}$&\\
&&1.9\% H$_{2}$&\\
\hline
300 K&68.6\% H$_{2}$&66.0\% H$_{2}$O&67.3\% CO$_{2}$\\
&8.4\% H$_{2}$O&6.6\% CH$_{4}$&15.6\% H$_{2}$O\\
&4.5\% CH$_{4}$&6.5\% N$_{2}$&12.8\% N$_{2}$\\
&1.2\% NH$_{3}$&4.9\% CO$_{2}$&\\
\hline
\multicolumn{4}{c}{The remaining gas is composed of He, bringing}\\
\multicolumn{4}{c}{totals to 100\%.}\\
\end{tabular}
\end{center}
\end{table}

\begin{figure}
\resizebox{5.5in}{!}
{\includegraphics{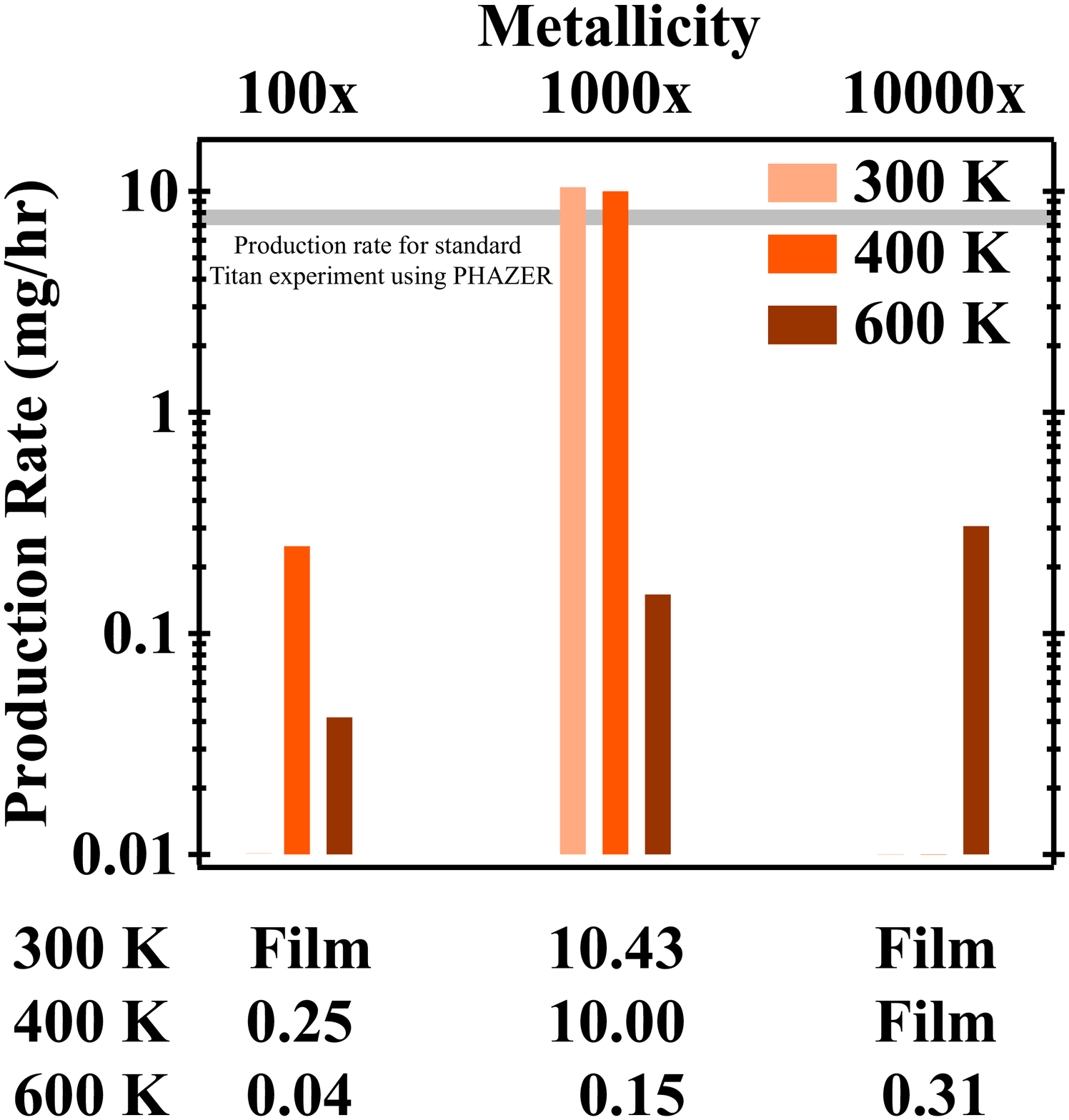}}
\caption{Shown here are the production rates measured for experiments shown in Figure \ref{fig:pies} and Table \ref{table:exp}. ``Film'' indicates that there was enough solid produced by the experiment to result in a visible film on the substrates, but there was not enough to collect and weigh. \label{fig:rate}}
\end{figure}

As shown in Figure \ref{fig:rate}, the H$_{2}$-dominated experiments (100x at 400 and 600 K; 1000x at 600 K) have lower haze production rates (0.04 and 0.25 mg/hr, respectively) than the other experiments with measurable production rates. This is not surprising given that previous works have shown that H$_{2}$ decreases particle formation in gas mixtures that include CH$_{4}$ and/or CO$_{2}$\cite{Raulin:1982, Dewitt:2009, Sciamma:2010} likely by termination of chains, which decreases production of larger molecules. However, the production rates are non-zero, indicating that these atmospheres may still be capable of producing a tenuous photochemically generated haze. The H$_{2}$-dominated giant planets in our own solar system have optically thin stratospheric hazes that are produced photochemically.

The two highest production rates measured ($\sim$10 mg/hr) were for the 1000x metallicity (H$_{2}$O-dominated) gas mixtures at 300 and 400 K. This production rate is actually higher than the production rate from our standard Titan experiment (5\% CH$_{4}$ in 95\% N$_{2}$), which produces $\sim$7.4 mg/hr using the identical setup\cite{He:2017}. The exoplanet experiments do have a higher CH$_{4}$ content than our standard Titan experiment, however they have significantly less N$_{2}$ and previous work has shown N$_{2}$ plays an important role in gas to particle conversion in N$_{2}$/CH$_{4}$ gas mixtures\cite{Imanaka:2010, Trainer:2012}. Additionally, previous work with plasma experiments indicates that addition of CO$_{2}$ should decrease particle formation\cite{Trainer:2004b}. On the other hand, previous work has shown that H$_{2}$O (present as a liquid and gas) may promote the formation of organics\cite{Miller:1953} so further work is necessary to understand how the complex interplay of these different chemical pathways is resulting in such a high production rate. 

The CO$_{2}$-dominated experiments (10000x) at 300 and 400 K did not produce sufficient sample to collect and weigh; however inspection of the quartz substrates reveals that a thin film is present indicating that solid was produced, albeit at a lower rate than the other experiments presented here. However, the 600 K experiment produced enough sample to collect. Aside from the increase in temperature, the main difference between this experiment and the 300 and 400 K experiments is the addition of CO and H$_{2}$ to the gas mixture. Given that H$_{2}$ is usually observed to decrease particle production as discussed above, it seems unlikely that the addition of H$_{2}$ to the gas mixture is responsible for the increase in production rate. However, previous work has shown that the addition of CO can increase the production rate\cite{Horst:2014}; CO is a better source of carbon for haze formation than CO$_{2}$, as dissociation of CO produces atomic carbon, which can be used to build larger organic molecules. The combination of CO and H$_{2}$O may be particularly efficient for increasing production rate as photochemical reactions  beginning with these two species produce a variety of organic compounds\cite{Bar-Nun:1983}.

The two highest production rate experiments have the two highest CH$_{4}$ concentrations, but the third highest production rate (10000x at 600 K) has no CH$_{4}$ at all, demonstrating that there are multiple pathways for organic haze formation and that CH$_{4}$ is not necessarily required. In that case the gas mixture had CO, which provided a source of carbon in place of CH$_{4}$. However, it is important to note that the production rates are not simply a function of carbon abundance, C/O, C/H, or C/N ratios in the initial gas mixtures. This result also demonstrates the need for experimental investigations to develop a robust theory of haze formation in planetary atmospheres. 

It is important to remember that the experimental matrix varied temperature and metallicity and that all 9 gas mixtures investigated are compositionally distinct. This approach makes interpreting the chemical pathways responsible for the observed production rates challenging, but provides our first insights into which regions of temperature and metallicity phase space may result in photochemical haze production, thereby enabling future investigations focused on specific pathways. The complexities of atmospheres compared to the laboratory makes it challenging to convert laboratory production rates into actual planetary haze production rates, but the relative comparisons between our experiments, in addition to the comparison to our nominal Titan rate, provide guidance on which gases are more likely to result in particle formation. 

Finally, we note that visual inspection of the films on the quartz substrates indicates large variation in particle color and efforts are underway to measure optical constants of these analog materials. 

We performed the first experimental simulations of atmospheric chemistry and haze formation relevant to super-Earth and mini-Neptune atmospheres. We find haze production rates that span at least two orders of magnitude. Two of the nine experiments (1000x solar metallicity at 300 and 400 K) yielded higher production rates than our standard Titan experiment and therefore likely represent conditions that are very favorable for haze formation. While further work is necessary to elucidate the chemical pathways responsible for these differences, particularly to understand the very high production rate cases, our experiments show that atmospheric characterization efforts for cool (T$<$800 K) super-Earth and mini-Neptune type exoplanets will encounter planets with a wide variety of haze production rates. These findings provide an importance balance in both caution and optimism for the planning of near-term observations of super-Earths and mini-Neptunes with facilities such as the James Webb Space Telescope.

\section{Methods} \label{sec:Methods}

\begin{figure}
\centering
\includegraphics{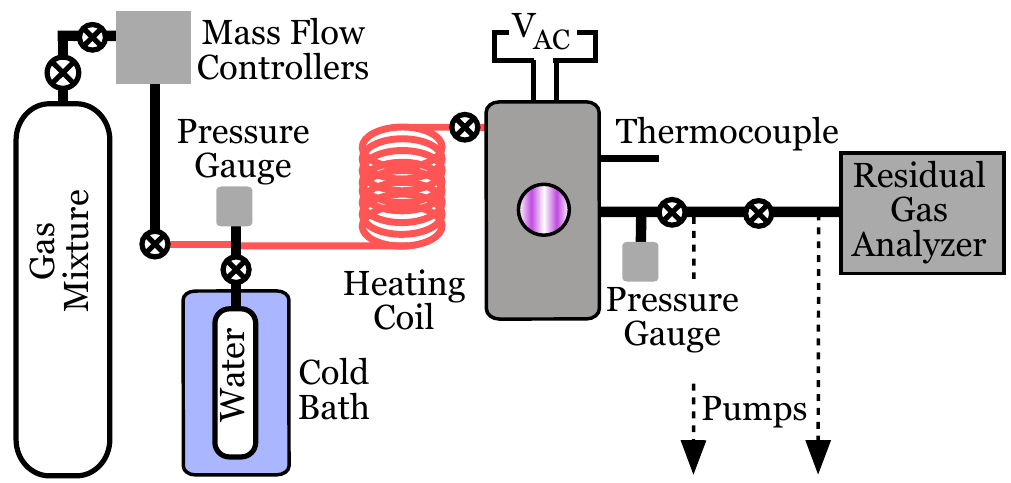}
\caption{Due to the large variety of gases used for these experiments, the schematic shown here provides a general idea of the setup. The details varied depending on the gases used with attention paid to solubility of gases in liquid water, condensation temperatures, and gas purity. \label{fig:exp}}
\end{figure}

We performed the experiments using the PHAZER (Planetary Haze Research) Chamber at Johns Hopkins University\cite{He:2017}. A schematic of the experimental setup is shown in Figure \ref{fig:exp}. We mix the reactant gases (CH$_{4}$ 99.999\%, CO 99.99\%, N$_{2}$ 99.9997\%, CO$_{2}$ 99.999\%, H$_{2}$ 99.9999\%, He 99.9995\% from Airgas) in a custom built mixing manifold designed to enable mixing multiple gases over a broad range of compositions. Since water is liquid at room temperature, it was introduced into the gas mixtures through the use of a dry ice temperature bath surrounding a canister of HPLC grade water (Fisher Chemical), which allowed for control of the water vapor pressure (and therefore resulting mixing ratio) by variation of the temperature of the bath (by adjusting the relative proportions of dry ice, methanol (CH$_{3}$OH), and water). For the one mixture that contained NH$_{3}$ (300 K, 100x), a solution of ammonium hydroxide (ACS grade, Millipore) was used with the dry ice bath to provide the NH$_{3}$ and H$_{2}$O required for the mixture. A liquid nitrogen cold trap was used to remove known contaminants from CO before mixing it with the other gases\cite{Horst:2014, He:2017}. The gas mixtures then flow through a custom heating coil, which allows temperature control from room temperature up to 800 K. The gases then flow continuously at a rate of 10 sccm (standard cubic centimeters per minute) through a stainless steel chamber using a mass flow controller (MKS Instruments, Inc.) maintaining a constant pressure of a few mbar (depending on the temperature) where they are exposed to the cold plasma generated by an AC glow discharge for approximately 3 seconds. The plasma produced by the AC glow discharge is used as a proxy for energetic processes occurring in planetary upper atmospheres. While it is not analogous to any specific process, we use it because it is sufficiently energetic to directly dissociate very stable molecules such as N$_{2}$ or CO, which is often accomplished in planetary atmospheres by EUV photons and therefore typically used as an analog for the relatively energetic environment of planetary upper atmospheres\cite{Cable:2012}. Note that the AC glow discharge is a cold plasma source and therefore the neutral gas temperature is not significantly altered by the plasma. The discharge dissociates and/or ionizes the reactant gases initiating chemical reactions in the chamber. Product gases and remaining reactant gases flow out of the chamber, while any solid produced in the experiment remains. Quartz substrates placed in the chamber collect thin films of particles and are used to check for particle production at very low production rates. The experiments run continuously for 72 hours and gas phase measurements demonstrate that the gas phase composition remains stable throughout this period of time after the initial equilibration time (less than 30 minutes)\cite{He:2017}. After 72 hours the plasma is shut off and the chamber is slowly returned to ambient temperature, while pumping to remove remaining gases. The chamber is placed in a dry ($<$0.1 ppm H$_{2}$O), oxygen free ($<$0.1 ppm O$_{2}$) N$_{2}$ glove box (Inert Technology Inc., I-lab 2GB) where they are removed from the chamber and weighed (Sartorius Entris 224-1S with standard deviation of 0.1 mg) providing a measurement of the production rate. 

\bibliography{titanoxygen}


\begin{addendum}
\item This work was supported by the NASA Exoplanets Research Program Grant NNX16AB45G. CH 
was supported by the Morton K. and Jane Blaustein Foundation.
\end{addendum}

\end{document}